\documentclass[aps,prl,twocolumn,superscriptaddress,groupedaddress]{revtex4}  
\usepackage{graphicx}  
\usepackage{dcolumn}   
\usepackage{bm}        
\usepackage{amssymb}   
\usepackage{siunitx}
\usepackage[breaklinks=true,colorlinks,citecolor=blue,linkcolor=blue,urlcolor=blue]{hyperref}
\usepackage{multirow}
\usepackage{xcolor}
\usepackage{amsmath}
\usepackage{hhline}
\usepackage{lipsum}

\usepackage{color} 
\newcommand{\Z}{\mathbb{Z} }

\begin{document}

\title{High Spin-Chern-Number Insulator in $\alpha$-Antimonene with a Hidden Topological Phase} 

\author{Baokai Wang}
\thanks{B. W. and X. Z. contributed equally to this work.}
\affiliation{Department of Physics, Northeastern University, Boston, Massachusetts 02115, USA}

\author{Xiaoting Zhou}
\thanks{B. W. and X. Z. contributed equally to this work.}
\affiliation{Department of Physics, Northeastern University, Boston, Massachusetts 02115, USA}

\author{Yen-Chuan Lin}
\affiliation{Department of Physics, National Taiwan University, Taipei 10617, Taiwan}

\author{Hsin Lin}
\email{nilnish@gmail.com}
\affiliation{Institute of Physics, Academia Sinica, Taipei 11529, Taiwan}

\author{Arun Bansil}
\email{ar.bansil@northeastern.edu}
\affiliation{Department of Physics, Northeastern University, Boston, Massachusetts 02115, USA}

\begin{abstract}
{In investigating the topological electronic structures of monolayer $\alpha$-phase group V elements, we uncover a new topological phase, which is invisible in the symmetry-based topological quantum chemistry (TQC) as well as symmetry indicators (SIs). Since $\alpha$ phase As and Sb share the same band representations at high-symmetry points, they are both trivial insulators in terms of TQC and SIs. We demonstrate, however, that there is a topological phase transition between As and Sb that involves a band-gap closing at two $k$-points on the high-symmetry $\rm{X}$-$\rm{\Gamma}$-$\rm{X}$ line. In the absence of spin-orbit coupling (SOC), As is a trivial insulator, while Sb is a Dirac semimetal with four Dirac points (DPs) located away from the high-symmetry lines. Inclusion of $S_z$-conserved SOC gaps out the Dirac points and induces a nontrivial Berry curvature and drives Sb into a high spin Chern number topological phase. The band structure of $\alpha$-Bi differs from that of Sb by a band inversion at $\Gamma$, transforming Bi into a $\Z_2$ topological insulator. Our study shows that quantized spin Hall conductivity can serve as a topological invariant beyond $\Z_2$ for characterizing topological phases.}

\end{abstract}

\maketitle

{\it Introduction.--} 
The past decade has witnessed remarkable advances in the field of topological materials\cite{TI_3, TI_1, TI_2, TI_4}. Novel bulk and boundary electronic properties of topological materials endow them with unique possibilities for fundamental studies as well as for applications. 
Topological physics took off with the discovery of the topological insulators (TIs) protected by time-reversal (TR) symmetry, which can be characterized by a $\Z_2$ index. Initially, the $\Z_2$ invariant was linked to the quantized spin Hall conductivity and expressed as $\Z_2$ = mod($\mathcal{C}_s$, 2) where $\mathcal{C}_s$ is the spin Chern number\cite{qsh_graphene, z2order, qsh_zhang, qsh_konig, nondissipative, spinchern}. It has been widely believed that $\Z_2$ and the spin Chern number $\mathcal{C}_s$ provide an equivalent description of a TR-invariant system. Accordingly, the two-dimensional (2D) TIs are also called quantum spin Hall insulators\cite{robustness, chernbi2se3, fukui}. We emphasize that while $\Z_2$ can only take two integral values (e.g. 0 and 1), this is not the case for the spin Chern number $\mathcal{C}_s$ in a quantum spin Hall insulator since $\mathcal{C}_s$ can, in principle, assume any integral value\cite{ezawahigh, trsbrokenqsh}. This immediately begs the question whether a $\Z_2$ trivial {\it topological} phase can be realized with a non-trivial spin Chern number. 

\begin{figure}
\includegraphics[width=\linewidth]{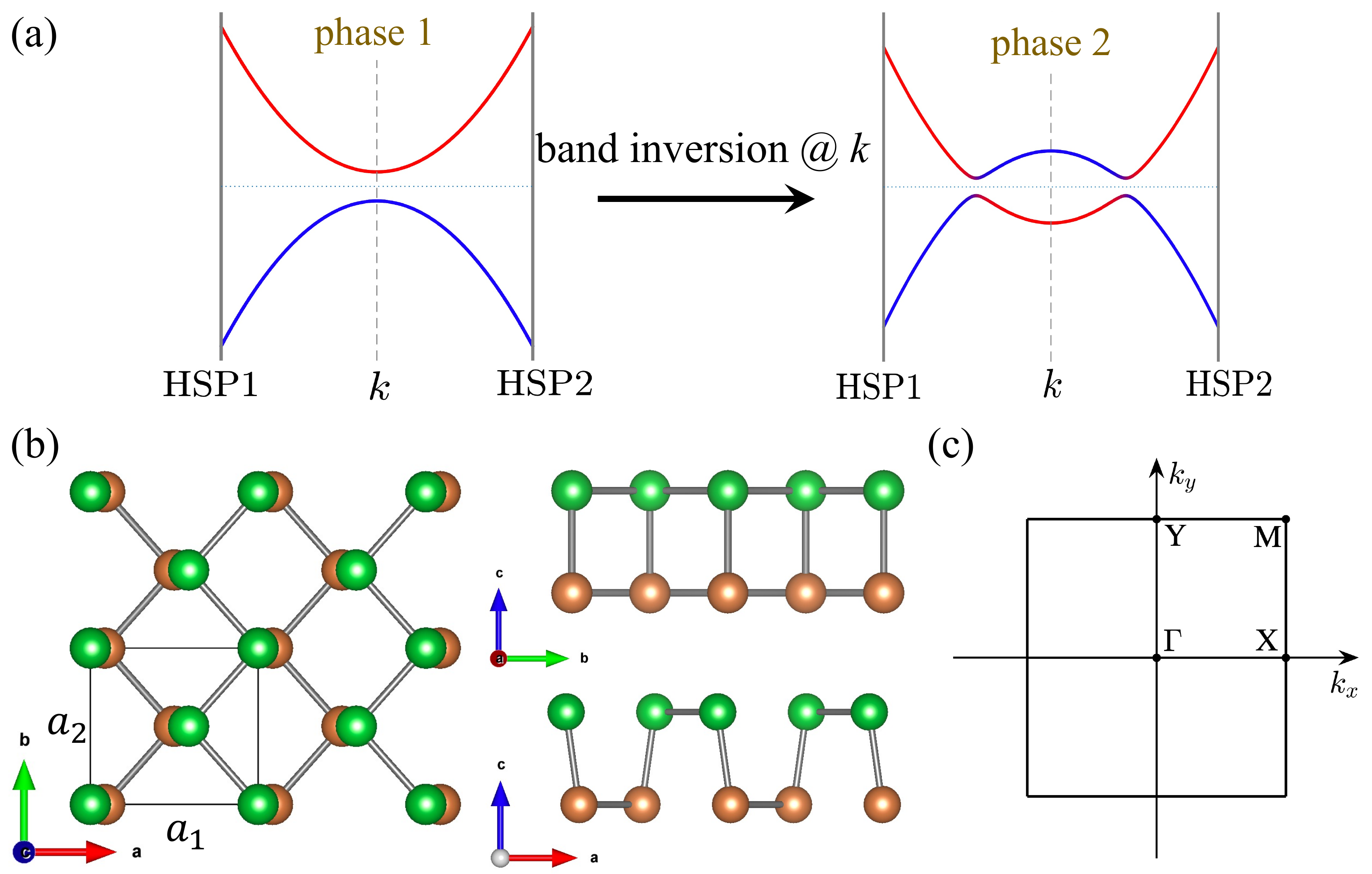}
\caption{
{(a) Illustration of a band inversion occurring at a generic $k$-point. (b) Crystal structure of $\alpha$-(As, Sb, Bi) in different views. (c) Brillouin zone of $\alpha$-(As, Sb, Bi).} 
}
\label{fig:1}
\end{figure}

Here, we discuss a topological phase characterized by a high spin Chern number in monolayer $\alpha$-Sb, which we show to be a trivial insulator when viewed through the lens of existing topological classification schemes\cite{tqc, si_1, si_2, si_3, si_4, combinatorics, mapping, catalogue, catalogue1}. In particular, within the framework of the TQC and SIs, by checking the band representations (BRs) at high-symmetry points (HSPs), one can diagnose the presence of stable, fragile topological, and atomic insulator phases through an analysis of band inversions at the HSPs in the Brillouin zone (BZ). This scheme, however, is $\it{incomplete}$ in the sense that it fails to recognize band inversions that occur at $k$-points other than the HSPs, see Fig.~\ref{fig:1}(a)).

We show that the electronic structures of monolayer $\alpha$-phase group V elements provide an exception to the TQC framework. As and Sb share the same band representations at HSPs and have both been classified as topologically trivial insulators using analysis based on TQC and SIs. However, we demonstrate that there is a phase transition in going from As to Sb. In the absence of SOC, As is an insulator but Sb is a Dirac semimetal. Band inversions between the two materials occur at two $k$-points on the high symmetry $\rm{X}$-$\Gamma$-$\rm{X}$ line in the BZ, resulting in the emergence of four Dirac points (DPs) in Sb. The inclusion of $S_z$-conserved SOC then gaps out the DPs, and induces a nontrivial Berry curvature around the DPs, which gives rise to a high spin Chern number $\mathcal{C}_s$ = 2. As a result, $\alpha$-Sb not only possesses larger spin Hall conductivity but it also harbors two pairs of gapless helical edge states, which drive a larger spin-polarized current. Notably, the band structure of Bi differs from that of Sb by an additional band inversion at the $\Gamma$ point, which drives Bi into becoming a quantum spin Hall insulator with $\Z_2 = 1$ as well as $\mathcal{C}_s=1$. The quantized spin Hall conductivity or spin Chern number can thus serve as a topological invariant beyond $\Z_2$ for characterizing a new type of topology in $\alpha$-phase group V elements.

 \begin{figure}
\includegraphics[width=\linewidth]{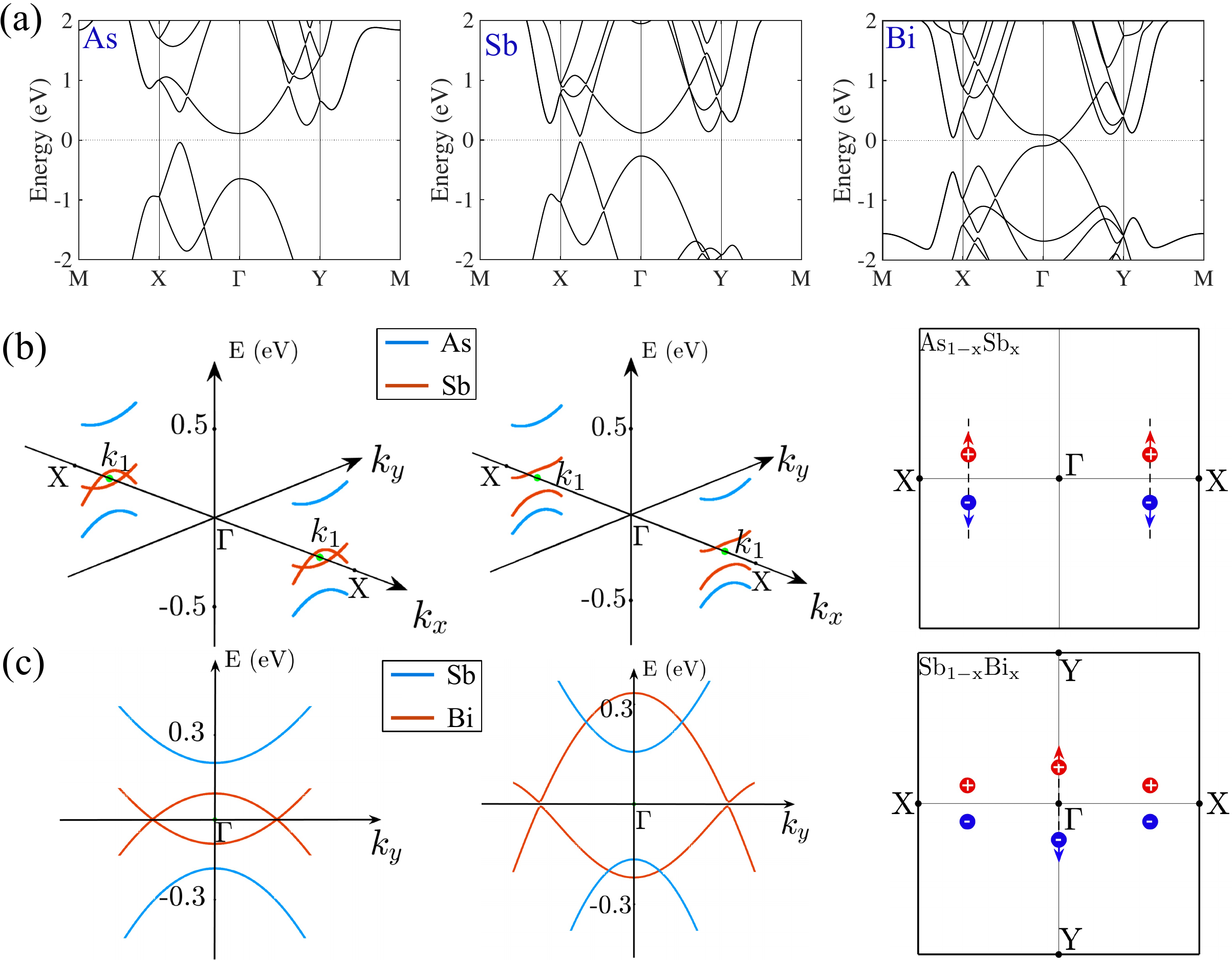}
\caption{
{(a) Band structure of $\alpha$-(As, Sb, Bi) in the absence of SOC. (b) Band inversions between As and Sb without (left panel) and with SOC (middle panel). The right panel illustrates the movement of Dirac points in $\rm{As}_{1-x}\rm{Sb}_x$ when $x$ exceeds 0.80. (c) Band inversions between Sb and Bi without (left panel) and with SOC (middle panel). The right panel illustrates the movement of Dirac points in $\rm{Sb}_{1-x}\rm{Bi}_x$ when $x$ exceeds 0.68.} 
}
\label{fig:2}
\end{figure}

{\it 2D bulk electronic structure--} Recently, $\alpha$-antimonene and  bithmuthene have been grown on a variety of substrates\cite{bismuthene, antimonene}. The monolayer $\alpha$-phase group V elements crystallize in space group $\mathcal{N}o. 53$ (layer group $\mathcal{N}o. 42$), which is generated by glide mirror $\mathcal{G} = \{M_z|\frac{1}{2}\frac{1}{2}0\}$, the inversion symmetry $\mathcal{P}$ and the mirror symmetry $M_y$. $\alpha$-(As, Sb, Bi) adopt a puckered structure.  It has two atomic planes (marked with green and orange colors, Fig.~\ref{fig:1}) having a vertical separation comparable to the bond length. In each atomic plane, the bonding between atoms forms zig-zag chains along the y-direction\cite{unpinned}. Throughout this paper, the electronic structures are calculated using realistic, material-specific  tight-binging models, which agree well with the corresponding first-principles calculations and capture the topology faithfully. Fig.~\ref{fig:2}(a) presents the band structures of $\alpha$-As, Sb, and Bi. The band structures around the Fermi level are dominated by the $p$-orbitals. In the absence of SOC, the band structure of $\rm{As}$ and $\rm{Sb}$ along the high-symmetry lines are well gapped, while for Bi, there is a band-crossing on the high-symmetry line $\rm{\Gamma}$-$\rm{Y}$, implying a band inversion from $\rm{Sb}$ to $\rm{Bi}$. A close examination of the band structures reveals that $\rm{Sb}$ supports four Dirac points at $k$-points close to the high-symmetry line $\rm{X}$-$\rm{\Gamma}$-$\rm{X}$. We study the evolution of the band structure of $\rm{As}_{1-x}\rm{Sb}_x$ as a function of the concentration $x$ of $\rm{Sb}$. When $x \approx 0.80$, the valence and conduction bands touch at $\bm{k}$ = $\pm 0.375 \rm{\bf{b_1}}$. As $x$ becomes larger, the touching-point splits into two parts, which move to the opposite sides of the $\rm{X}$-$\rm{\Gamma}$-$\rm{X}$ line, as depicted in Fig.~\ref{fig:2}(b). The two touching-points carry opposite chirality. For $\rm{Sb}_{1-x}Bi_x$, as the concentration of $\rm{Bi}$ increases to 0.68, a band inversion occurs at $\Gamma$. As $x$ becomes larger, the two Dirac points appear on $\rm{\Gamma}$-$\rm{Y}$, which finally result in 6 Dirac nodes in the BZ for $\rm{Bi}$, as depicted in Fig.~\ref{fig:2}(c). Fig.~\ref{fig:2}(b, c) illustrates the two types of band inversions that  occur between (As, Sb) and (Sb, Bi), respectively. The band inversion between As and Sb happens at the two $k_1$ points (Fig.~\ref{fig:2}(b)), but does not alter the symmetry eigenvalues at HSPs, and $\rm{Sb}$ and $\rm{As}$ admit the same band representations. Therefore, these two materials are classified into the same category according to TQC. In contrast, the band inversion between Sb and Bi occurs at $\Gamma$, and since the band representations of Sb and Bi are different at $\Gamma$, Sb and Bi can be expected to belong to different topological phases.

We now consider the effects of adding spin-orbit coupling that preserves the $S_z$ component. In the presence of SOC, the band structure of $\rm{As}$ remains almost invariant due to the negligible strength of SOC. For $\rm{Sb}$ and $\rm{Bi}$, the SOC gaps out all the Dirac cones, leading to a well-gapped band structure. Through an analysis of the parities of valence bands, we find that As and Sb are trivial insulators with $\Z_2 = 0$ while Bi is a topological insulator with $\Z_2=1$.
 
\begin{figure}
\includegraphics[width=\linewidth]{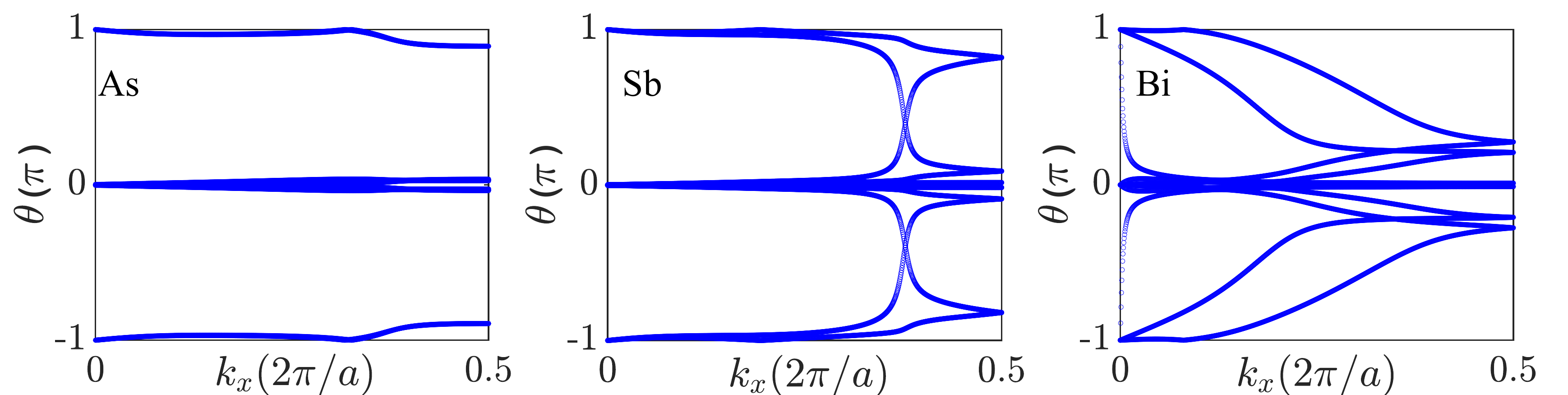}
\caption{
{(a-c) Spectrum of Wannier Charge Centers in (a) As, (b) Sb and (c) Bi. } 
}
\label{fig:2_2}
\end{figure}

As illustrated in Figure \ref{fig:2}(b), a band inversion happens between As and Sb, indicating that the two materials may be topologically distinct. In order to gain further insight, we plot the spectrum of Wannier charge centers (WCCs) in Fig.~\ref{fig:2_2}. The three materials are seen to display very different patterns of WCCs. For As, the spectrum has a large gap, while both Sb and Bi show gapless features in WCCs, implying they are topologically nontrivial. An essential difference between the patters of WCCs of Sb and Bi is that the $\Z_2$ number read from WCCs in Sb is 0 while it is 1 for Bi. Therefore, another topological invariant is needed to fully characterize the phase in Sb, which is invisible to the $\Z_2$ invariant.

{\it band representations--} According to TQC, topological phases of well-gapped materials can be diagnosed by band representations (BRs). 
We give the BRs of As and Sb at the HSPs in Table.~\ref{tab1}. It will be seen that As and Sb have the same band representations at HSPs and can be decomposed into a linear combination of elementary band representations (EBRs). This agrees well with the analysis of valence band parities, confirming that both As and Sb are trivial insulators. Note, however, that although the BRs of As and Sb can be decomposed into linear combinations of EBRs,  these cannot be expressed as linear combinations of atomic band representations (aBRs). Therefore, they belong to obstructed atomic insulators\cite{unconventional, fillingenforced}. Consider $\rm{As}_{1-x}\rm{Sb}_x$ with the doping from $x=1$ to $x=0$, equivalent to annihilating the two pairs of DPs in Sb, it can easily reach the conclusion that As is an obstructed atomic insulator according to the reference\cite{topo_obstructed}.

\begin{table}
\small
\caption{Irreducible representations of valence bands in $\alpha$-(As, Sb).}
\begin{centering}
\scalebox{1.0}{
\begin{tabular}{c|c|c|c|c}
\hline\hline
  \#53 & $\Gamma$ & X & Y  & M 
 \cr\hline
 $\alpha$-(As, Sb)& $\Gamma$1+(1) &X1(2)  & Y1(2) & M1+(2) \\
                 & $\Gamma$4-(1)& X2(2) & Y2(2) & M1-(2)\\
                 &  $\Gamma$1-(1) & X1(2) & Y1(2) & M1-(2)\\
                 &  $\Gamma$1+(1) &	&	 &  \\
                 &  $\Gamma$4+(1) &	&	 & \\
                 &  $\Gamma$3+(1) &	&	 &
\cr\hline\hline 
\end{tabular}}
\end{centering}
\label{tab1}
\end{table}

\begin{figure}
\includegraphics[width=\linewidth]{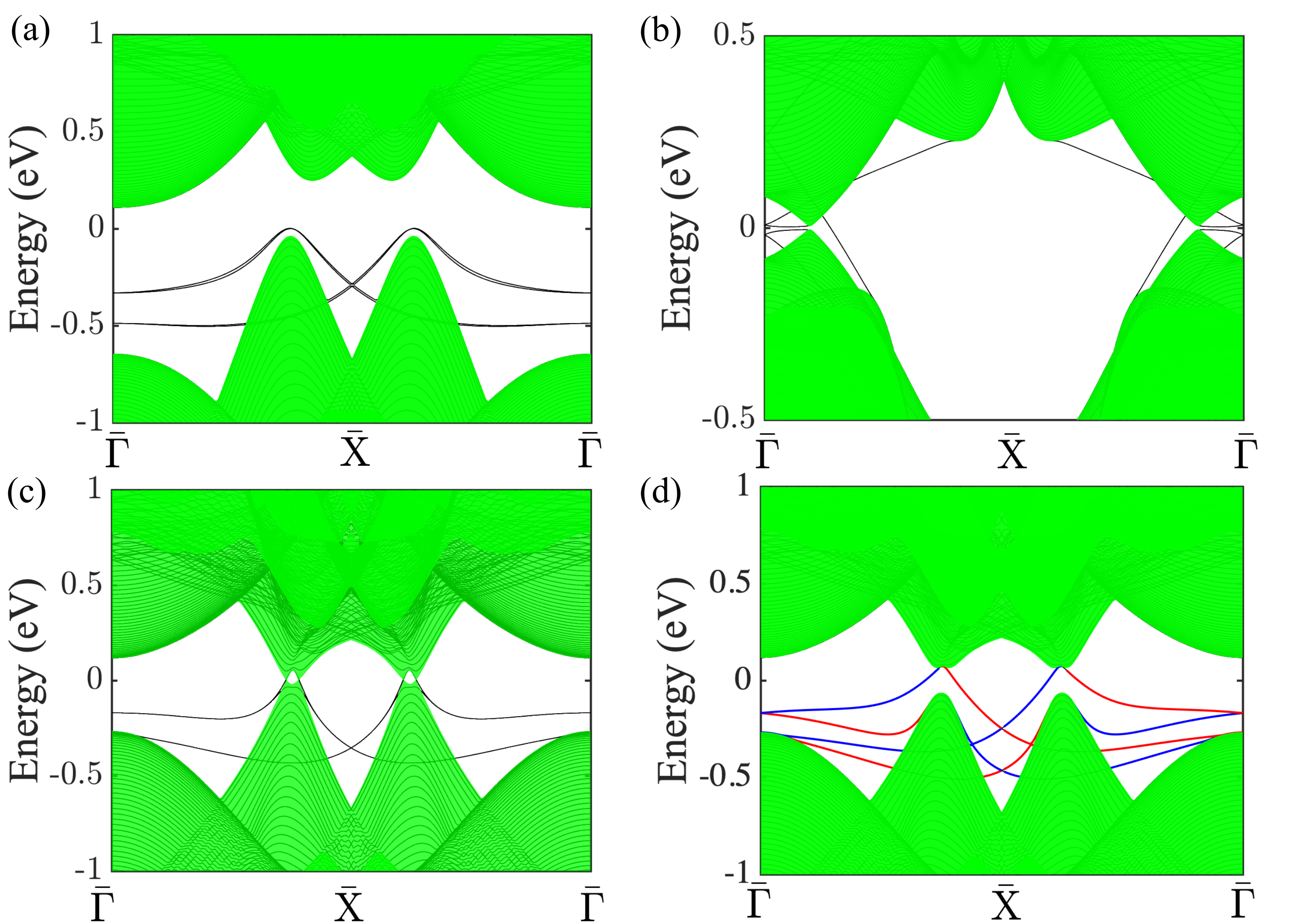}
\caption{
{\bf{Band structures of nanoribbon along $\bf{a_1 + a_2}$}}. (a, b) As and Bi in the presence of SOC. (c, d) Sb in the absence or presence of SOC. The red and blue colored bands denote the spin-up and spin-down states, respectively.
}
\label{fig:3}
\end{figure}

{\it Nanoribbon band structures--} We consider a nanoribbon extending along the direction $\bf{a_1 + a_2}$, which preserves the glide symmetry  $\{M_z|\frac{1}{2}\frac{1}{2}0\}$. The band structures are presented in Fig.~\ref{fig:3}. We can see that As has a gapped band structure and supports a group of hourglass fermions due to the glide symmetry\cite{hourglass}. For Bi, the gapless edge states transverse the bulk band gap, displaying the typical zigzag connectivity signaling a strong topological insulator phase. The edge bands of Sb display more interesting features. When the SOC is excluded, there are four DPs in the BZ (see Fig.~\ref{fig:2}(b)). Fermi arcs connect the DPs with opposite chiralities in the ribbon spectrum, see Fig.~\ref{fig:3}(c). The inclusion of SOC gaps out the DPs and splits the Fermi arcs into two pairs of gapless edge bands across the bulk band gap. Moreover, the edge bands are fully spin-polarized as marked by different colors in Fig. ~\ref{fig:3}(d). We also examined the edge band structures of $\hat{x}$ or $\hat{y}$ directed ribbons, which are also found to support spin-polarized gapless edge states. Hence it confirms that this distinct topological phase is a pure bulk property and obeys bulk-edge correspondence. The spin-polarized gapless edge state provides another signature of  the topological phase in Sb. It gives us a hint that the spin Hall conductivity could be a topological invariant to characterize the topological phase in the $\alpha$-phase group V elements.

It is interesting to compare the electronic structures of $\alpha$-Sb and graphene, which possess close resemblance\cite{qsh_graphene}. In the absence of SOC, both materials are Dirac semimetals with the Fermi arcs appearing on the edges\cite{fermiarc}. The inclusion of SOC gaps the DPs and splits the Fermi arc states into gapless edge states in both materials. Sb can be viewed as hosting two copies of the gapless edge states of graphene. However, we emphasize that there is an essential difference between Sb and graphene. The Dirac nodes are located at generic $k$-points in Sb, while these are located at K and K' points in graphene. The band inversions leave Sb in the topologically trivial phase in the sense of $\Z_2=0$ but turn graphene into a topological insulator.

\begin{figure}
\includegraphics[width=\linewidth]{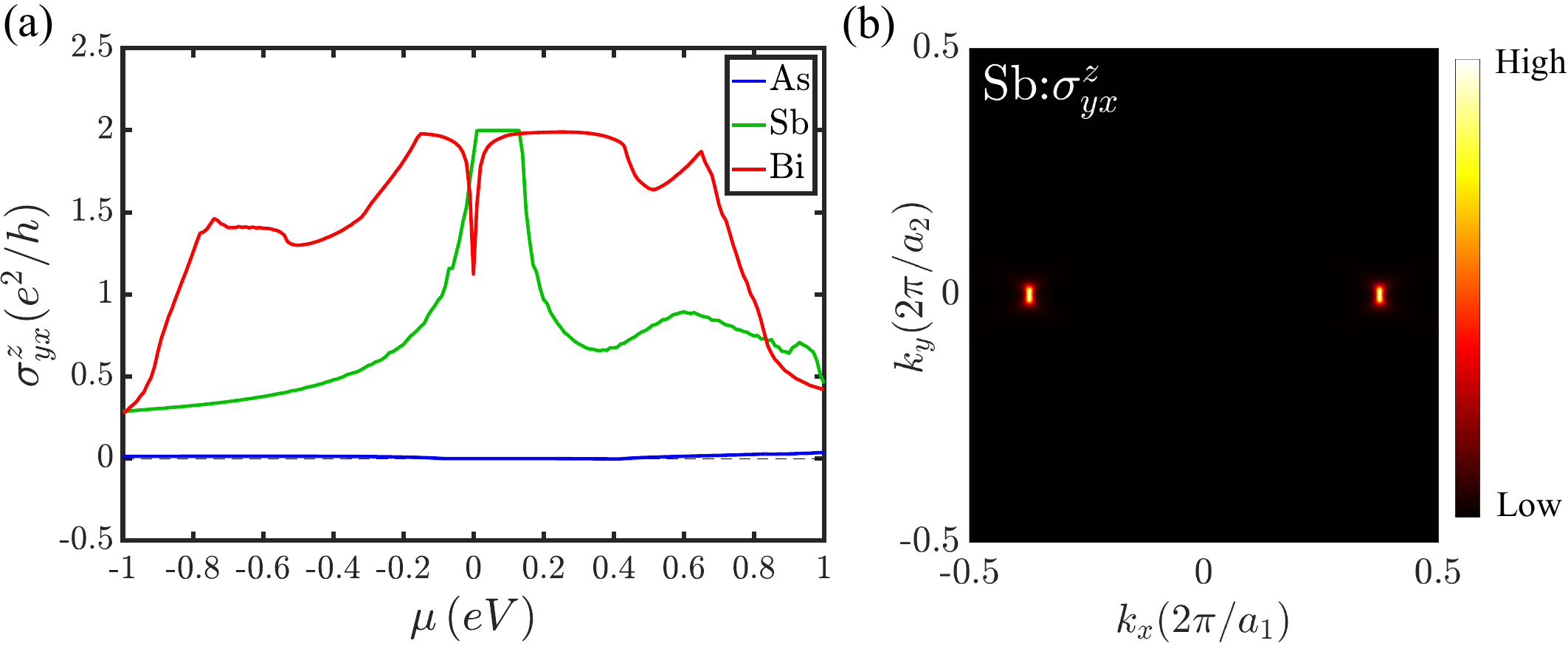}
\caption{
{(a) Spin Hall conductivity of $\alpha$-(As, Sb, Bi) marked by different colors. (b) The distribution of spin Berry curvature in Brillouin zone for $\alpha$-Sb.}}
\label{fig:4}
\end{figure}

{\it Quantized Spin Hall conductivity--}
As discussed in the preceding section, Sb hosts spin-polarized gapless edge states indicating that it can support nontrivial spin transport. Accordingly, we turn now to consider spin Hall conductivity. Fig.~\ref{fig:4}(a) presents the $\sigma_{yx}^{z}$ at different chemical potentials. The spin Hall conductivity is zero for As, as expected, since the nanoribbon has a gapped band structure. In contrast, the spin Hall conductivity of Sb shows a quantized plateau $\sigma_{yx}^{z} = 2\cdot \frac{e^2}{h}$ inside the bulk band gap, i.e. $\mathcal{C}_s = 2$.  This high quantized spin Hall conductivity distinguishes  Sb from As and makes Sb topologically distinct, although the two materials share the same band representations at HSPs. We plot the spin Berry curvature of Sb in the BZ in Fig.~\ref{fig:4}(b). It can be seen that the spin Berry curvature is mainly distributed around the DPs and vanishes elsewhere. Bi also features a quantized spin Hall conductivity with $\mathcal{C}_s = 1$ that agrees with the above analysis of $\Z_2$, which can also be expressed as $mod(\mathcal{C}_s, 2)$.

 
{\it Discussion and Conclusion.--}
In Sb, the bands with $S_z$-conserved SOC can be smoothly deformed into the bands with full SOC without closing the band gap or breaking a crystal symmetry. Since the SOC generally breaks the conservation of spin components in materials, $\sigma_{yx}^{z}$ is no longer quantized. However, our analysis shows that it still takes a nearly quantized value even in the presence of full SOC with deviation from $2\frac{e^2}{h}$ of less than 1\%. Therefore, Sb provides a good candidate for experimental realization of our predicted novel quantum spin Hall effect. 

Interestingly, our study continues the long history of Bi and Sb in inspiring advances in the field of topological materials. This is perhaps not surprising. The first 3D $\Z_2$ topological insulator was realized in Bi/Sb alloys\cite{hsieh}. 3D Bi is $\Z_2$ trivial, and Sb is $\Z_2$ nontrivial. Both end-compounds are semimetals, and an insulating band gap can be found in their alloy with $\Z_2 = 1$. An early tight-binding model of Bi/Sb led to the consideration of mirror Chern numbers\cite{teo}, although first-principles calculations predict Bi to be a trivial mirror Chern insulator. Bi was considered to be topologically trivial until it was identified as a high-order topological insulator \cite{highorder}. A further analysis of Bi based on the SIs found it to host a first-order topological phase with gapless surface states protected by a rotational symmetry\cite{bi_tci}. The surface Dirac cones in Bi are not located at high symmetry points but at generic $k$-points in the surface Brillouin zone, making it difficult to the associated nontrivial topologies. 2D monolayer $\alpha$-Sb in our study not only exhibits edge-state related Dirac cones at generic $k$-points but also hosts band inversions away from HSPs, so that both TQC and SIs fail to detect its nontrivial topology. On the one hand, we uncovered a novel topological phase in monolayer $\alpha$-Sb, which can be characterized by a high spin Chern number $\mathcal{C}_s = 2$ but is $\Z_2$ trivial. It exemplifies the distinction between $\Z_2$ topological insulators and quantum spin Hall insulators. Our work demonstrates that the $\Z_2$ topological insulator phase and quantum spin Hall insulator belong to different categories in the zoo of topological phases. On the other hand, our study reveals the importance of band inversions at generic $k$-point, which do not affect the band representations at HSPs but can lead to distinct topological phases. It emphasizes the importance of examining wavefunction in the entire BZ to fully characterize the topological phases rather than only considering the wavefunctions at HSPs \cite{topology_trs, 1dtmi, beyondtqc}.

Our study reveals that even `trivial' insulators can host quantized conductivities and opens the door for the discovery of new classes of topological materials, which lie outside the scope of analysis based on s TQC and SIs. It thus dramatically expands the possibilities for candidate materials for spintronics and quantum information sciences applications.

{\it Acknowledgement. } H.L. acknowledges the support by the Ministry of Science and Technology (MOST) in Taiwan under grant number MOST109-2112-M-001-014-MY3. The work at Northeastern University is supported by the Air Force Office of Scientific Research under award number FA9550-20-1-0322, and it benefited from computational resources of Northeastern University's Advanced Scientific Computation Center (ASCC) and the Discovery Cluster.

{\it Notes added}: This work was submitted on Dec 2nd, 2021, and is under review.  

\bibliographystyle{apsrev4-1}

\begin{thebibliography}{39}%
\makeatletter
\providecommand \@ifxundefined [1]{%
 \@ifx{#1\undefined}
}%
\providecommand \@ifnum [1]{%
 \ifnum #1\expandafter \@firstoftwo
 \else \expandafter \@secondoftwo
 \fi
}%
\providecommand \@ifx [1]{%
 \ifx #1\expandafter \@firstoftwo
 \else \expandafter \@secondoftwo
 \fi
}%
\providecommand \natexlab [1]{#1}%
\providecommand \enquote  [1]{``#1''}%
\providecommand \bibnamefont  [1]{#1}%
\providecommand \bibfnamefont [1]{#1}%
\providecommand \citenamefont [1]{#1}%
\providecommand \href@noop [0]{\@secondoftwo}%
\providecommand \href [0]{\begingroup \@sanitize@url \@href}%
\providecommand \@href[1]{\@@startlink{#1}\@@href}%
\providecommand \@@href[1]{\endgroup#1\@@endlink}%
\providecommand \@sanitize@url [0]{\catcode `\\12\catcode `\$12\catcode
  `\&12\catcode `\#12\catcode `\^12\catcode `\_12\catcode `\%12\relax}%
\providecommand \@@startlink[1]{}%
\providecommand \@@endlink[0]{}%
\providecommand \url  [0]{\begingroup\@sanitize@url \@url }%
\providecommand \@url [1]{\endgroup\@href {#1}{\urlprefix }}%
\providecommand \urlprefix  [0]{URL }%
\providecommand \Eprint [0]{\href }%
\providecommand \doibase [0]{http://dx.doi.org/}%
\providecommand \selectlanguage [0]{\@gobble}%
\providecommand \bibinfo  [0]{\@secondoftwo}%
\providecommand \bibfield  [0]{\@secondoftwo}%
\providecommand \translation [1]{[#1]}%
\providecommand \BibitemOpen [0]{}%
\providecommand \bibitemStop [0]{}%
\providecommand \bibitemNoStop [0]{.\EOS\space}%
\providecommand \EOS [0]{\spacefactor3000\relax}%
\providecommand \BibitemShut  [1]{\csname bibitem#1\endcsname}%
\let\auto@bib@innerbib\@empty
\bibitem [{\citenamefont {Bansil}\ \emph {et~al.}(2016)\citenamefont {Bansil},
  \citenamefont {Lin},\ and\ \citenamefont {Das}}]{TI_3}%
  \BibitemOpen
  \bibfield  {author} {\bibinfo {author} {\bibfnamefont {A.}~\bibnamefont
  {Bansil}}, \bibinfo {author} {\bibfnamefont {H.}~\bibnamefont {Lin}}, \ and\
  \bibinfo {author} {\bibfnamefont {T.}~\bibnamefont {Das}},\ }\href {\doibase
  10.1103/RevModPhys.88.021004} {\bibfield  {journal} {\bibinfo  {journal}
  {Rev. Mod. Phys.}\ }\textbf {\bibinfo {volume} {88}},\ \bibinfo {pages}
  {021004} (\bibinfo {year} {2016})}\BibitemShut {NoStop}%
\bibitem [{\citenamefont {Hasan}\ and\ \citenamefont {Kane}(2010)}]{TI_1}%
  \BibitemOpen
  \bibfield  {author} {\bibinfo {author} {\bibfnamefont {M.~Z.}\ \bibnamefont
  {Hasan}}\ and\ \bibinfo {author} {\bibfnamefont {C.~L.}\ \bibnamefont
  {Kane}},\ }\href {\doibase 10.1103/RevModPhys.82.3045} {\bibfield  {journal}
  {\bibinfo  {journal} {Rev. Mod. Phys.}\ }\textbf {\bibinfo {volume} {82}},\
  \bibinfo {pages} {3045} (\bibinfo {year} {2010})}\BibitemShut {NoStop}%
\bibitem [{\citenamefont {Qi}\ and\ \citenamefont {Zhang}(2011)}]{TI_2}%
  \BibitemOpen
  \bibfield  {author} {\bibinfo {author} {\bibfnamefont {X.-L.}\ \bibnamefont
  {Qi}}\ and\ \bibinfo {author} {\bibfnamefont {S.-C.}\ \bibnamefont {Zhang}},\
  }\href {\doibase 10.1103/RevModPhys.83.1057} {\bibfield  {journal} {\bibinfo
  {journal} {Rev. Mod. Phys.}\ }\textbf {\bibinfo {volume} {83}},\ \bibinfo
  {pages} {1057} (\bibinfo {year} {2011})}\BibitemShut {NoStop}%
\bibitem [{\citenamefont {Armitage}\ \emph {et~al.}(2018)\citenamefont
  {Armitage}, \citenamefont {Mele},\ and\ \citenamefont {Vishwanath}}]{TI_4}%
  \BibitemOpen
  \bibfield  {author} {\bibinfo {author} {\bibfnamefont {N.~P.}\ \bibnamefont
  {Armitage}}, \bibinfo {author} {\bibfnamefont {E.~J.}\ \bibnamefont {Mele}},
  \ and\ \bibinfo {author} {\bibfnamefont {A.}~\bibnamefont {Vishwanath}},\
  }\href {\doibase 10.1103/RevModPhys.90.015001} {\bibfield  {journal}
  {\bibinfo  {journal} {Rev. Mod. Phys.}\ }\textbf {\bibinfo {volume} {90}},\
  \bibinfo {pages} {015001} (\bibinfo {year} {2018})}\BibitemShut {NoStop}%
\bibitem [{\citenamefont {Kane}\ and\ \citenamefont
  {Mele}(2005{\natexlab{a}})}]{qsh_graphene}%
  \BibitemOpen
  \bibfield  {author} {\bibinfo {author} {\bibfnamefont {C.~L.}\ \bibnamefont
  {Kane}}\ and\ \bibinfo {author} {\bibfnamefont {E.~J.}\ \bibnamefont
  {Mele}},\ }\href {\doibase 10.1103/PhysRevLett.95.226801} {\bibfield
  {journal} {\bibinfo  {journal} {Phys. Rev. Lett.}\ }\textbf {\bibinfo
  {volume} {95}},\ \bibinfo {pages} {226801} (\bibinfo {year}
  {2005}{\natexlab{a}})}\BibitemShut {NoStop}%
\bibitem [{\citenamefont {Kane}\ and\ \citenamefont
  {Mele}(2005{\natexlab{b}})}]{z2order}%
  \BibitemOpen
  \bibfield  {author} {\bibinfo {author} {\bibfnamefont {C.~L.}\ \bibnamefont
  {Kane}}\ and\ \bibinfo {author} {\bibfnamefont {E.~J.}\ \bibnamefont
  {Mele}},\ }\href {\doibase 10.1103/PhysRevLett.95.146802} {\bibfield
  {journal} {\bibinfo  {journal} {Phys. Rev. Lett.}\ }\textbf {\bibinfo
  {volume} {95}},\ \bibinfo {pages} {146802} (\bibinfo {year}
  {2005}{\natexlab{b}})}\BibitemShut {NoStop}%
\bibitem [{\citenamefont {Bernevig}\ \emph {et~al.}(2006)\citenamefont
  {Bernevig}, \citenamefont {Hughes},\ and\ \citenamefont {Zhang}}]{qsh_zhang}%
  \BibitemOpen
  \bibfield  {author} {\bibinfo {author} {\bibfnamefont {B.~A.}\ \bibnamefont
  {Bernevig}}, \bibinfo {author} {\bibfnamefont {T.~L.}\ \bibnamefont
  {Hughes}}, \ and\ \bibinfo {author} {\bibfnamefont {S.-C.}\ \bibnamefont
  {Zhang}},\ }\href {\doibase 10.1126/science.1133734} {\bibfield  {journal}
  {\bibinfo  {journal} {Science}\ }\textbf {\bibinfo {volume} {314}},\ \bibinfo
  {pages} {1757} (\bibinfo {year} {2006})}\BibitemShut {NoStop}%
\bibitem [{\citenamefont {König}\ \emph {et~al.}(2007)\citenamefont {König},
  \citenamefont {Wiedmann}, \citenamefont {Brüne}, \citenamefont {Roth},
  \citenamefont {Buhmann}, \citenamefont {Molenkamp}, \citenamefont {Qi},\ and\
  \citenamefont {Zhang}}]{qsh_konig}%
  \BibitemOpen
  \bibfield  {author} {\bibinfo {author} {\bibfnamefont {M.}~\bibnamefont
  {König}}, \bibinfo {author} {\bibfnamefont {S.}~\bibnamefont {Wiedmann}},
  \bibinfo {author} {\bibfnamefont {C.}~\bibnamefont {Brüne}}, \bibinfo
  {author} {\bibfnamefont {A.}~\bibnamefont {Roth}}, \bibinfo {author}
  {\bibfnamefont {H.}~\bibnamefont {Buhmann}}, \bibinfo {author} {\bibfnamefont
  {L.~W.}\ \bibnamefont {Molenkamp}}, \bibinfo {author} {\bibfnamefont {X.-L.}\
  \bibnamefont {Qi}}, \ and\ \bibinfo {author} {\bibfnamefont {S.-C.}\
  \bibnamefont {Zhang}},\ }\href {\doibase 10.1126/science.1148047} {\bibfield
  {journal} {\bibinfo  {journal} {Science}\ }\textbf {\bibinfo {volume}
  {318}},\ \bibinfo {pages} {766} (\bibinfo {year} {2007})}\BibitemShut
  {NoStop}%
\bibitem [{\citenamefont {Sheng}\ \emph {et~al.}(2005)\citenamefont {Sheng},
  \citenamefont {Sheng}, \citenamefont {Ting},\ and\ \citenamefont
  {Haldane}}]{nondissipative}%
  \BibitemOpen
  \bibfield  {author} {\bibinfo {author} {\bibfnamefont {L.}~\bibnamefont
  {Sheng}}, \bibinfo {author} {\bibfnamefont {D.~N.}\ \bibnamefont {Sheng}},
  \bibinfo {author} {\bibfnamefont {C.~S.}\ \bibnamefont {Ting}}, \ and\
  \bibinfo {author} {\bibfnamefont {F.~D.~M.}\ \bibnamefont {Haldane}},\ }\href
  {\doibase 10.1103/PhysRevLett.95.136602} {\bibfield  {journal} {\bibinfo
  {journal} {Phys. Rev. Lett.}\ }\textbf {\bibinfo {volume} {95}},\ \bibinfo
  {pages} {136602} (\bibinfo {year} {2005})}\BibitemShut {NoStop}%
\bibitem [{\citenamefont {Sheng}\ \emph {et~al.}(2006)\citenamefont {Sheng},
  \citenamefont {Weng}, \citenamefont {Sheng},\ and\ \citenamefont
  {Haldane}}]{spinchern}%
  \BibitemOpen
  \bibfield  {author} {\bibinfo {author} {\bibfnamefont {D.~N.}\ \bibnamefont
  {Sheng}}, \bibinfo {author} {\bibfnamefont {Z.~Y.}\ \bibnamefont {Weng}},
  \bibinfo {author} {\bibfnamefont {L.}~\bibnamefont {Sheng}}, \ and\ \bibinfo
  {author} {\bibfnamefont {F.~D.~M.}\ \bibnamefont {Haldane}},\ }\href
  {\doibase 10.1103/PhysRevLett.97.036808} {\bibfield  {journal} {\bibinfo
  {journal} {Phys. Rev. Lett.}\ }\textbf {\bibinfo {volume} {97}},\ \bibinfo
  {pages} {036808} (\bibinfo {year} {2006})}\BibitemShut {NoStop}%
\bibitem [{\citenamefont {Prodan}(2009)}]{robustness}%
  \BibitemOpen
  \bibfield  {author} {\bibinfo {author} {\bibfnamefont {E.}~\bibnamefont
  {Prodan}},\ }\href {\doibase 10.1103/PhysRevB.80.125327} {\bibfield
  {journal} {\bibinfo  {journal} {Phys. Rev. B}\ }\textbf {\bibinfo {volume}
  {80}},\ \bibinfo {pages} {125327} (\bibinfo {year} {2009})}\BibitemShut
  {NoStop}%
\bibitem [{\citenamefont {Li}\ \emph {et~al.}(2010)\citenamefont {Li},
  \citenamefont {Sheng}, \citenamefont {Sheng},\ and\ \citenamefont
  {Xing}}]{chernbi2se3}%
  \BibitemOpen
  \bibfield  {author} {\bibinfo {author} {\bibfnamefont {H.}~\bibnamefont
  {Li}}, \bibinfo {author} {\bibfnamefont {L.}~\bibnamefont {Sheng}}, \bibinfo
  {author} {\bibfnamefont {D.~N.}\ \bibnamefont {Sheng}}, \ and\ \bibinfo
  {author} {\bibfnamefont {D.~Y.}\ \bibnamefont {Xing}},\ }\href {\doibase
  10.1103/PhysRevB.82.165104} {\bibfield  {journal} {\bibinfo  {journal} {Phys.
  Rev. B}\ }\textbf {\bibinfo {volume} {82}},\ \bibinfo {pages} {165104}
  (\bibinfo {year} {2010})}\BibitemShut {NoStop}%
\bibitem [{\citenamefont {Fukui}\ and\ \citenamefont {Hatsugai}(2007)}]{fukui}%
  \BibitemOpen
  \bibfield  {author} {\bibinfo {author} {\bibfnamefont {T.}~\bibnamefont
  {Fukui}}\ and\ \bibinfo {author} {\bibfnamefont {Y.}~\bibnamefont
  {Hatsugai}},\ }\href {\doibase 10.1103/PhysRevB.75.121403} {\bibfield
  {journal} {\bibinfo  {journal} {Phys. Rev. B}\ }\textbf {\bibinfo {volume}
  {75}},\ \bibinfo {pages} {121403} (\bibinfo {year} {2007})}\BibitemShut
  {NoStop}%
\bibitem [{\citenamefont {Ezawa}(2013)}]{ezawahigh}%
  \BibitemOpen
  \bibfield  {author} {\bibinfo {author} {\bibfnamefont {M.}~\bibnamefont
  {Ezawa}},\ }\href {\doibase 10.1038/srep03435} {\bibfield  {journal}
  {\bibinfo  {journal} {Scientific Reports}\ }\textbf {\bibinfo {volume} {3}},\
  \bibinfo {pages} {3435} (\bibinfo {year} {2013})}\BibitemShut {NoStop}%
\bibitem [{\citenamefont {Yang}\ \emph {et~al.}(2011)\citenamefont {Yang},
  \citenamefont {Xu}, \citenamefont {Sheng}, \citenamefont {Wang},
  \citenamefont {Xing},\ and\ \citenamefont {Sheng}}]{trsbrokenqsh}%
  \BibitemOpen
  \bibfield  {author} {\bibinfo {author} {\bibfnamefont {Y.}~\bibnamefont
  {Yang}}, \bibinfo {author} {\bibfnamefont {Z.}~\bibnamefont {Xu}}, \bibinfo
  {author} {\bibfnamefont {L.}~\bibnamefont {Sheng}}, \bibinfo {author}
  {\bibfnamefont {B.}~\bibnamefont {Wang}}, \bibinfo {author} {\bibfnamefont
  {D.~Y.}\ \bibnamefont {Xing}}, \ and\ \bibinfo {author} {\bibfnamefont
  {D.~N.}\ \bibnamefont {Sheng}},\ }\href {\doibase
  10.1103/PhysRevLett.107.066602} {\bibfield  {journal} {\bibinfo  {journal}
  {Phys. Rev. Lett.}\ }\textbf {\bibinfo {volume} {107}},\ \bibinfo {pages}
  {066602} (\bibinfo {year} {2011})}\BibitemShut {NoStop}%
\bibitem [{\citenamefont {Bradlyn}\ \emph {et~al.}(2017)\citenamefont
  {Bradlyn}, \citenamefont {Elcoro}, \citenamefont {Cano}, \citenamefont
  {Vergniory}, \citenamefont {Wang}, \citenamefont {Felser}, \citenamefont
  {Aroyo},\ and\ \citenamefont {Bernevig}}]{tqc}%
  \BibitemOpen
  \bibfield  {author} {\bibinfo {author} {\bibfnamefont {B.}~\bibnamefont
  {Bradlyn}}, \bibinfo {author} {\bibfnamefont {L.}~\bibnamefont {Elcoro}},
  \bibinfo {author} {\bibfnamefont {J.}~\bibnamefont {Cano}}, \bibinfo {author}
  {\bibfnamefont {M.~G.}\ \bibnamefont {Vergniory}}, \bibinfo {author}
  {\bibfnamefont {Z.}~\bibnamefont {Wang}}, \bibinfo {author} {\bibfnamefont
  {C.}~\bibnamefont {Felser}}, \bibinfo {author} {\bibfnamefont {M.~I.}\
  \bibnamefont {Aroyo}}, \ and\ \bibinfo {author} {\bibfnamefont {B.~A.}\
  \bibnamefont {Bernevig}},\ }\href {\doibase 10.1038/nature23268} {\bibfield
  {journal} {\bibinfo  {journal} {Nature}\ }\textbf {\bibinfo {volume} {547}},\
  \bibinfo {pages} {298} (\bibinfo {year} {2017})}\BibitemShut {NoStop}%
\bibitem [{\citenamefont {Po}\ \emph {et~al.}(2017)\citenamefont {Po},
  \citenamefont {Vishwanath},\ and\ \citenamefont {Watanabe}}]{si_1}%
  \BibitemOpen
  \bibfield  {author} {\bibinfo {author} {\bibfnamefont {H.~C.}\ \bibnamefont
  {Po}}, \bibinfo {author} {\bibfnamefont {A.}~\bibnamefont {Vishwanath}}, \
  and\ \bibinfo {author} {\bibfnamefont {H.}~\bibnamefont {Watanabe}},\ }\href
  {\doibase 10.1038/s41467-017-00133-2} {\bibfield  {journal} {\bibinfo
  {journal} {Nature Communications}\ }\textbf {\bibinfo {volume} {8}},\
  \bibinfo {pages} {50} (\bibinfo {year} {2017})}\BibitemShut {NoStop}%
\bibitem [{\citenamefont {Tang}\ \emph
  {et~al.}(2019{\natexlab{a}})\citenamefont {Tang}, \citenamefont {Po},
  \citenamefont {Vishwanath},\ and\ \citenamefont {Wan}}]{si_2}%
  \BibitemOpen
  \bibfield  {author} {\bibinfo {author} {\bibfnamefont {F.}~\bibnamefont
  {Tang}}, \bibinfo {author} {\bibfnamefont {H.~C.}\ \bibnamefont {Po}},
  \bibinfo {author} {\bibfnamefont {A.}~\bibnamefont {Vishwanath}}, \ and\
  \bibinfo {author} {\bibfnamefont {X.}~\bibnamefont {Wan}},\ }\href {\doibase
  10.1038/s41586-019-0937-5} {\bibfield  {journal} {\bibinfo  {journal}
  {Nature}\ }\textbf {\bibinfo {volume} {566}},\ \bibinfo {pages} {486}
  (\bibinfo {year} {2019}{\natexlab{a}})}\BibitemShut {NoStop}%
\bibitem [{\citenamefont {Tang}\ \emph
  {et~al.}(2019{\natexlab{b}})\citenamefont {Tang}, \citenamefont {Po},
  \citenamefont {Vishwanath},\ and\ \citenamefont {Wan}}]{si_3}%
  \BibitemOpen
  \bibfield  {author} {\bibinfo {author} {\bibfnamefont {F.}~\bibnamefont
  {Tang}}, \bibinfo {author} {\bibfnamefont {H.~C.}\ \bibnamefont {Po}},
  \bibinfo {author} {\bibfnamefont {A.}~\bibnamefont {Vishwanath}}, \ and\
  \bibinfo {author} {\bibfnamefont {X.}~\bibnamefont {Wan}},\ }\href {\doibase
  10.1038/s41567-019-0418-7} {\bibfield  {journal} {\bibinfo  {journal} {Nature
  Physics}\ }\textbf {\bibinfo {volume} {15}},\ \bibinfo {pages} {470}
  (\bibinfo {year} {2019}{\natexlab{b}})}\BibitemShut {NoStop}%
\bibitem [{\citenamefont {Tang}\ \emph
  {et~al.}(2019{\natexlab{c}})\citenamefont {Tang}, \citenamefont {Po},
  \citenamefont {Vishwanath},\ and\ \citenamefont {Wan}}]{si_4}%
  \BibitemOpen
  \bibfield  {author} {\bibinfo {author} {\bibfnamefont {F.}~\bibnamefont
  {Tang}}, \bibinfo {author} {\bibfnamefont {H.~C.}\ \bibnamefont {Po}},
  \bibinfo {author} {\bibfnamefont {A.}~\bibnamefont {Vishwanath}}, \ and\
  \bibinfo {author} {\bibfnamefont {X.}~\bibnamefont {Wan}},\ }\href {\doibase
  10.1126/sciadv.aau8725} {\bibfield  {journal} {\bibinfo  {journal} {Science
  Advances}\ }\textbf {\bibinfo {volume} {5}} (\bibinfo {year}
  {2019}{\natexlab{c}}),\ 10.1126/sciadv.aau8725},\ \Eprint
  {http://arxiv.org/abs/https://advances.sciencemag.org/content/5/3/eaau8725.full.pdf}
  {https://advances.sciencemag.org/content/5/3/eaau8725.full.pdf} \BibitemShut
  {NoStop}%
\bibitem [{\citenamefont {Kruthoff}\ \emph {et~al.}(2017)\citenamefont
  {Kruthoff}, \citenamefont {de~Boer}, \citenamefont {van Wezel}, \citenamefont
  {Kane},\ and\ \citenamefont {Slager}}]{combinatorics}%
  \BibitemOpen
  \bibfield  {author} {\bibinfo {author} {\bibfnamefont {J.}~\bibnamefont
  {Kruthoff}}, \bibinfo {author} {\bibfnamefont {J.}~\bibnamefont {de~Boer}},
  \bibinfo {author} {\bibfnamefont {J.}~\bibnamefont {van Wezel}}, \bibinfo
  {author} {\bibfnamefont {C.~L.}\ \bibnamefont {Kane}}, \ and\ \bibinfo
  {author} {\bibfnamefont {R.-J.}\ \bibnamefont {Slager}},\ }\href {\doibase
  10.1103/PhysRevX.7.041069} {\bibfield  {journal} {\bibinfo  {journal} {Phys.
  Rev. X}\ }\textbf {\bibinfo {volume} {7}},\ \bibinfo {pages} {041069}
  (\bibinfo {year} {2017})}\BibitemShut {NoStop}%
\bibitem [{\citenamefont {Song}\ \emph {et~al.}(2018)\citenamefont {Song},
  \citenamefont {Zhang}, \citenamefont {Fang},\ and\ \citenamefont
  {Fang}}]{mapping}%
  \BibitemOpen
  \bibfield  {author} {\bibinfo {author} {\bibfnamefont {Z.}~\bibnamefont
  {Song}}, \bibinfo {author} {\bibfnamefont {T.}~\bibnamefont {Zhang}},
  \bibinfo {author} {\bibfnamefont {Z.}~\bibnamefont {Fang}}, \ and\ \bibinfo
  {author} {\bibfnamefont {C.}~\bibnamefont {Fang}},\ }\href {\doibase
  10.1038/s41467-018-06010-w} {\bibfield  {journal} {\bibinfo  {journal}
  {Nature Communications}\ }\textbf {\bibinfo {volume} {9}},\ \bibinfo {pages}
  {3530} (\bibinfo {year} {2018})}\BibitemShut {NoStop}%
\bibitem [{\citenamefont {Zhang}\ \emph {et~al.}(2019)\citenamefont {Zhang},
  \citenamefont {Jiang}, \citenamefont {Song}, \citenamefont {Huang},
  \citenamefont {He}, \citenamefont {Fang}, \citenamefont {Weng},\ and\
  \citenamefont {Fang}}]{catalogue}%
  \BibitemOpen
  \bibfield  {author} {\bibinfo {author} {\bibfnamefont {T.}~\bibnamefont
  {Zhang}}, \bibinfo {author} {\bibfnamefont {Y.}~\bibnamefont {Jiang}},
  \bibinfo {author} {\bibfnamefont {Z.}~\bibnamefont {Song}}, \bibinfo {author}
  {\bibfnamefont {H.}~\bibnamefont {Huang}}, \bibinfo {author} {\bibfnamefont
  {Y.}~\bibnamefont {He}}, \bibinfo {author} {\bibfnamefont {Z.}~\bibnamefont
  {Fang}}, \bibinfo {author} {\bibfnamefont {H.}~\bibnamefont {Weng}}, \ and\
  \bibinfo {author} {\bibfnamefont {C.}~\bibnamefont {Fang}},\ }\href {\doibase
  10.1038/s41586-019-0944-6} {\bibfield  {journal} {\bibinfo  {journal}
  {Nature}\ }\textbf {\bibinfo {volume} {566}},\ \bibinfo {pages} {475}
  (\bibinfo {year} {2019})}\BibitemShut {NoStop}%
\bibitem [{\citenamefont {Vergniory}\ \emph {et~al.}(2019)\citenamefont
  {Vergniory}, \citenamefont {Elcoro}, \citenamefont {Felser}, \citenamefont
  {Regnault}, \citenamefont {Bernevig},\ and\ \citenamefont
  {Wang}}]{catalogue1}%
  \BibitemOpen
  \bibfield  {author} {\bibinfo {author} {\bibfnamefont {M.~G.}\ \bibnamefont
  {Vergniory}}, \bibinfo {author} {\bibfnamefont {L.}~\bibnamefont {Elcoro}},
  \bibinfo {author} {\bibfnamefont {C.}~\bibnamefont {Felser}}, \bibinfo
  {author} {\bibfnamefont {N.}~\bibnamefont {Regnault}}, \bibinfo {author}
  {\bibfnamefont {B.~A.}\ \bibnamefont {Bernevig}}, \ and\ \bibinfo {author}
  {\bibfnamefont {Z.}~\bibnamefont {Wang}},\ }\href {\doibase
  10.1038/s41586-019-0954-4} {\bibfield  {journal} {\bibinfo  {journal}
  {Nature}\ }\textbf {\bibinfo {volume} {566}},\ \bibinfo {pages} {480}
  (\bibinfo {year} {2019})}\BibitemShut {NoStop}%
\bibitem [{\citenamefont {Kowalczyk}\ \emph {et~al.}(2020)\citenamefont
  {Kowalczyk}, \citenamefont {Brown}, \citenamefont {Maerkl}, \citenamefont
  {Lu}, \citenamefont {Chiu}, \citenamefont {Liu}, \citenamefont {Yang},
  \citenamefont {Wang}, \citenamefont {Zasada}, \citenamefont {Genuzio},
  \citenamefont {Menteş}, \citenamefont {Locatelli}, \citenamefont {Chiang},\
  and\ \citenamefont {Bian}}]{bismuthene}%
  \BibitemOpen
  \bibfield  {author} {\bibinfo {author} {\bibfnamefont {P.~J.}\ \bibnamefont
  {Kowalczyk}}, \bibinfo {author} {\bibfnamefont {S.~A.}\ \bibnamefont
  {Brown}}, \bibinfo {author} {\bibfnamefont {T.}~\bibnamefont {Maerkl}},
  \bibinfo {author} {\bibfnamefont {Q.}~\bibnamefont {Lu}}, \bibinfo {author}
  {\bibfnamefont {C.-K.}\ \bibnamefont {Chiu}}, \bibinfo {author}
  {\bibfnamefont {Y.}~\bibnamefont {Liu}}, \bibinfo {author} {\bibfnamefont
  {S.~A.}\ \bibnamefont {Yang}}, \bibinfo {author} {\bibfnamefont
  {X.}~\bibnamefont {Wang}}, \bibinfo {author} {\bibfnamefont {I.}~\bibnamefont
  {Zasada}}, \bibinfo {author} {\bibfnamefont {F.}~\bibnamefont {Genuzio}},
  \bibinfo {author} {\bibfnamefont {T.~O.}\ \bibnamefont {Menteş}}, \bibinfo
  {author} {\bibfnamefont {A.}~\bibnamefont {Locatelli}}, \bibinfo {author}
  {\bibfnamefont {T.-C.}\ \bibnamefont {Chiang}}, \ and\ \bibinfo {author}
  {\bibfnamefont {G.}~\bibnamefont {Bian}},\ }\href {\doibase
  10.1021/acsnano.9b08136} {\bibfield  {journal} {\bibinfo  {journal} {ACS
  Nano}\ }\textbf {\bibinfo {volume} {14}},\ \bibinfo {pages} {1888} (\bibinfo
  {year} {2020})},\ \bibinfo {note} {pMID: 31971774},\ \Eprint
  {http://arxiv.org/abs/https://doi.org/10.1021/acsnano.9b08136}
  {https://doi.org/10.1021/acsnano.9b08136} \BibitemShut {NoStop}%
\bibitem [{\citenamefont {Lu}\ \emph {et~al.}(2021)\citenamefont {Lu},
  \citenamefont {Chen}, \citenamefont {Snyder}, \citenamefont {Cook},
  \citenamefont {Nguyen}, \citenamefont {Reddy}, \citenamefont {Chang},\ and\
  \citenamefont {Bian}}]{antimonene}%
  \BibitemOpen
  \bibfield  {author} {\bibinfo {author} {\bibfnamefont {Q.}~\bibnamefont
  {Lu}}, \bibinfo {author} {\bibfnamefont {K.~Y.}\ \bibnamefont {Chen}},
  \bibinfo {author} {\bibfnamefont {M.}~\bibnamefont {Snyder}}, \bibinfo
  {author} {\bibfnamefont {J.}~\bibnamefont {Cook}}, \bibinfo {author}
  {\bibfnamefont {D.~T.}\ \bibnamefont {Nguyen}}, \bibinfo {author}
  {\bibfnamefont {P.~V.~S.}\ \bibnamefont {Reddy}}, \bibinfo {author}
  {\bibfnamefont {T.-R.}\ \bibnamefont {Chang}}, \ and\ \bibinfo {author}
  {\bibfnamefont {G.}~\bibnamefont {Bian}},\ }\href@noop {} {\enquote {\bibinfo
  {title} {Observation of symmetry-protected dirac states in nonsymmorphic
  $\alpha$-antimonene},}\ } (\bibinfo {year} {2021}),\ \Eprint
  {http://arxiv.org/abs/2101.05793} {arXiv:2101.05793 [cond-mat.mes-hall]}
  \BibitemShut {NoStop}%
\bibitem [{\citenamefont {Lu}\ \emph {et~al.}(2016)\citenamefont {Lu},
  \citenamefont {Zhou}, \citenamefont {Chang}, \citenamefont {Guan},
  \citenamefont {Chen}, \citenamefont {Jiang}, \citenamefont {Jiang},
  \citenamefont {Wang}, \citenamefont {Yang}, \citenamefont {Feng},
  \citenamefont {Kawazoe},\ and\ \citenamefont {Lin}}]{unpinned}%
  \BibitemOpen
  \bibfield  {author} {\bibinfo {author} {\bibfnamefont {Y.}~\bibnamefont
  {Lu}}, \bibinfo {author} {\bibfnamefont {D.}~\bibnamefont {Zhou}}, \bibinfo
  {author} {\bibfnamefont {G.}~\bibnamefont {Chang}}, \bibinfo {author}
  {\bibfnamefont {S.}~\bibnamefont {Guan}}, \bibinfo {author} {\bibfnamefont
  {W.}~\bibnamefont {Chen}}, \bibinfo {author} {\bibfnamefont {Y.}~\bibnamefont
  {Jiang}}, \bibinfo {author} {\bibfnamefont {J.}~\bibnamefont {Jiang}},
  \bibinfo {author} {\bibfnamefont {X.-s.}\ \bibnamefont {Wang}}, \bibinfo
  {author} {\bibfnamefont {S.~A.}\ \bibnamefont {Yang}}, \bibinfo {author}
  {\bibfnamefont {Y.~P.}\ \bibnamefont {Feng}}, \bibinfo {author}
  {\bibfnamefont {Y.}~\bibnamefont {Kawazoe}}, \ and\ \bibinfo {author}
  {\bibfnamefont {H.}~\bibnamefont {Lin}},\ }\href {\doibase
  10.1038/npjcompumats.2016.11} {\bibfield  {journal} {\bibinfo  {journal} {npj
  Computational Materials}\ }\textbf {\bibinfo {volume} {2}},\ \bibinfo {pages}
  {16011} (\bibinfo {year} {2016})}\BibitemShut {NoStop}%
\bibitem [{\citenamefont {Gao}\ \emph {et~al.}(2021)\citenamefont {Gao},
  \citenamefont {Qian}, \citenamefont {Jia}, \citenamefont {Guo}, \citenamefont
  {Fang}, \citenamefont {Liu}, \citenamefont {Weng},\ and\ \citenamefont
  {Wang}}]{unconventional}%
  \BibitemOpen
  \bibfield  {author} {\bibinfo {author} {\bibfnamefont {J.}~\bibnamefont
  {Gao}}, \bibinfo {author} {\bibfnamefont {Y.}~\bibnamefont {Qian}}, \bibinfo
  {author} {\bibfnamefont {H.}~\bibnamefont {Jia}}, \bibinfo {author}
  {\bibfnamefont {Z.}~\bibnamefont {Guo}}, \bibinfo {author} {\bibfnamefont
  {Z.}~\bibnamefont {Fang}}, \bibinfo {author} {\bibfnamefont {M.}~\bibnamefont
  {Liu}}, \bibinfo {author} {\bibfnamefont {H.}~\bibnamefont {Weng}}, \ and\
  \bibinfo {author} {\bibfnamefont {Z.}~\bibnamefont {Wang}},\ }\href@noop {}
  {\enquote {\bibinfo {title} {Unconventional materials: the mismatch between
  electronic charge centers andatomic positions},}\ } (\bibinfo {year}
  {2021}),\ \Eprint {http://arxiv.org/abs/2106.08035} {arXiv:2106.08035
  [cond-mat.mtrl-sci]} \BibitemShut {NoStop}%
\bibitem [{\citenamefont {Xu}\ \emph {et~al.}(2021)\citenamefont {Xu},
  \citenamefont {Elcoro}, \citenamefont {Song}, \citenamefont {Vergniory},
  \citenamefont {Felser}, \citenamefont {Parkin}, \citenamefont {Regnault},
  \citenamefont {Mañes},\ and\ \citenamefont {Bernevig}}]{fillingenforced}%
  \BibitemOpen
  \bibfield  {author} {\bibinfo {author} {\bibfnamefont {Y.}~\bibnamefont
  {Xu}}, \bibinfo {author} {\bibfnamefont {L.}~\bibnamefont {Elcoro}}, \bibinfo
  {author} {\bibfnamefont {Z.-D.}\ \bibnamefont {Song}}, \bibinfo {author}
  {\bibfnamefont {M.~G.}\ \bibnamefont {Vergniory}}, \bibinfo {author}
  {\bibfnamefont {C.}~\bibnamefont {Felser}}, \bibinfo {author} {\bibfnamefont
  {S.~S.~P.}\ \bibnamefont {Parkin}}, \bibinfo {author} {\bibfnamefont
  {N.}~\bibnamefont {Regnault}}, \bibinfo {author} {\bibfnamefont {J.~L.}\
  \bibnamefont {Mañes}}, \ and\ \bibinfo {author} {\bibfnamefont {B.~A.}\
  \bibnamefont {Bernevig}},\ }\href@noop {} {\enquote {\bibinfo {title}
  {Filling-enforced obstructed atomic insulators},}\ } (\bibinfo {year}
  {2021}),\ \Eprint {http://arxiv.org/abs/2106.10276} {arXiv:2106.10276
  [cond-mat.mtrl-sci]} \BibitemShut {NoStop}%
\bibitem [{\citenamefont {Radha}\ and\ \citenamefont
  {Lambrecht}(2021)}]{topo_obstructed}%
  \BibitemOpen
  \bibfield  {author} {\bibinfo {author} {\bibfnamefont {S.~K.}\ \bibnamefont
  {Radha}}\ and\ \bibinfo {author} {\bibfnamefont {W.~R.~L.}\ \bibnamefont
  {Lambrecht}},\ }\href {\doibase 10.1103/PhysRevB.103.075435} {\bibfield
  {journal} {\bibinfo  {journal} {Phys. Rev. B}\ }\textbf {\bibinfo {volume}
  {103}},\ \bibinfo {pages} {075435} (\bibinfo {year} {2021})}\BibitemShut
  {NoStop}%
\bibitem [{\citenamefont {Wang}\ \emph {et~al.}(2016)\citenamefont {Wang},
  \citenamefont {Alexandradinata}, \citenamefont {Cava},\ and\ \citenamefont
  {Bernevig}}]{hourglass}%
  \BibitemOpen
  \bibfield  {author} {\bibinfo {author} {\bibfnamefont {Z.}~\bibnamefont
  {Wang}}, \bibinfo {author} {\bibfnamefont {A.}~\bibnamefont
  {Alexandradinata}}, \bibinfo {author} {\bibfnamefont {R.~J.}\ \bibnamefont
  {Cava}}, \ and\ \bibinfo {author} {\bibfnamefont {B.~A.}\ \bibnamefont
  {Bernevig}},\ }\href {\doibase 10.1038/nature17410} {\bibfield  {journal}
  {\bibinfo  {journal} {Nature}\ }\textbf {\bibinfo {volume} {532}},\ \bibinfo
  {pages} {189} (\bibinfo {year} {2016})}\BibitemShut {NoStop}%
\bibitem [{\citenamefont {Wan}\ \emph {et~al.}(2011)\citenamefont {Wan},
  \citenamefont {Turner}, \citenamefont {Vishwanath},\ and\ \citenamefont
  {Savrasov}}]{fermiarc}%
  \BibitemOpen
  \bibfield  {author} {\bibinfo {author} {\bibfnamefont {X.}~\bibnamefont
  {Wan}}, \bibinfo {author} {\bibfnamefont {A.~M.}\ \bibnamefont {Turner}},
  \bibinfo {author} {\bibfnamefont {A.}~\bibnamefont {Vishwanath}}, \ and\
  \bibinfo {author} {\bibfnamefont {S.~Y.}\ \bibnamefont {Savrasov}},\ }\href
  {\doibase 10.1103/PhysRevB.83.205101} {\bibfield  {journal} {\bibinfo
  {journal} {Phys. Rev. B}\ }\textbf {\bibinfo {volume} {83}},\ \bibinfo
  {pages} {205101} (\bibinfo {year} {2011})}\BibitemShut {NoStop}%
\bibitem [{\citenamefont {Hsieh}\ \emph {et~al.}(2008)\citenamefont {Hsieh},
  \citenamefont {Qian}, \citenamefont {Wray}, \citenamefont {Xia},
  \citenamefont {Hor}, \citenamefont {Cava},\ and\ \citenamefont
  {Hasan}}]{hsieh}%
  \BibitemOpen
  \bibfield  {author} {\bibinfo {author} {\bibfnamefont {D.}~\bibnamefont
  {Hsieh}}, \bibinfo {author} {\bibfnamefont {D.}~\bibnamefont {Qian}},
  \bibinfo {author} {\bibfnamefont {L.}~\bibnamefont {Wray}}, \bibinfo {author}
  {\bibfnamefont {Y.}~\bibnamefont {Xia}}, \bibinfo {author} {\bibfnamefont
  {Y.~S.}\ \bibnamefont {Hor}}, \bibinfo {author} {\bibfnamefont {R.~J.}\
  \bibnamefont {Cava}}, \ and\ \bibinfo {author} {\bibfnamefont {M.~Z.}\
  \bibnamefont {Hasan}},\ }\href {\doibase 10.1038/nature06843} {\bibfield
  {journal} {\bibinfo  {journal} {Nature}\ }\textbf {\bibinfo {volume} {452}},\
  \bibinfo {pages} {970} (\bibinfo {year} {2008})}\BibitemShut {NoStop}%
\bibitem [{\citenamefont {Teo}\ \emph {et~al.}(2008)\citenamefont {Teo},
  \citenamefont {Fu},\ and\ \citenamefont {Kane}}]{teo}%
  \BibitemOpen
  \bibfield  {author} {\bibinfo {author} {\bibfnamefont {J.~C.~Y.}\
  \bibnamefont {Teo}}, \bibinfo {author} {\bibfnamefont {L.}~\bibnamefont
  {Fu}}, \ and\ \bibinfo {author} {\bibfnamefont {C.~L.}\ \bibnamefont
  {Kane}},\ }\href {\doibase 10.1103/PhysRevB.78.045426} {\bibfield  {journal}
  {\bibinfo  {journal} {Phys. Rev. B}\ }\textbf {\bibinfo {volume} {78}},\
  \bibinfo {pages} {045426} (\bibinfo {year} {2008})}\BibitemShut {NoStop}%
\bibitem [{\citenamefont {Schindler}\ \emph {et~al.}(2018)\citenamefont
  {Schindler}, \citenamefont {Wang}, \citenamefont {Vergniory}, \citenamefont
  {Cook}, \citenamefont {Murani}, \citenamefont {Sengupta}, \citenamefont
  {Kasumov}, \citenamefont {Deblock}, \citenamefont {Jeon}, \citenamefont
  {Drozdov}, \citenamefont {Bouchiat}, \citenamefont {Gu{\'e}ron},
  \citenamefont {Yazdani}, \citenamefont {Bernevig},\ and\ \citenamefont
  {Neupert}}]{highorder}%
  \BibitemOpen
  \bibfield  {author} {\bibinfo {author} {\bibfnamefont {F.}~\bibnamefont
  {Schindler}}, \bibinfo {author} {\bibfnamefont {Z.}~\bibnamefont {Wang}},
  \bibinfo {author} {\bibfnamefont {M.~G.}\ \bibnamefont {Vergniory}}, \bibinfo
  {author} {\bibfnamefont {A.~M.}\ \bibnamefont {Cook}}, \bibinfo {author}
  {\bibfnamefont {A.}~\bibnamefont {Murani}}, \bibinfo {author} {\bibfnamefont
  {S.}~\bibnamefont {Sengupta}}, \bibinfo {author} {\bibfnamefont {A.~Y.}\
  \bibnamefont {Kasumov}}, \bibinfo {author} {\bibfnamefont {R.}~\bibnamefont
  {Deblock}}, \bibinfo {author} {\bibfnamefont {S.}~\bibnamefont {Jeon}},
  \bibinfo {author} {\bibfnamefont {I.}~\bibnamefont {Drozdov}}, \bibinfo
  {author} {\bibfnamefont {H.}~\bibnamefont {Bouchiat}}, \bibinfo {author}
  {\bibfnamefont {S.}~\bibnamefont {Gu{\'e}ron}}, \bibinfo {author}
  {\bibfnamefont {A.}~\bibnamefont {Yazdani}}, \bibinfo {author} {\bibfnamefont
  {B.~A.}\ \bibnamefont {Bernevig}}, \ and\ \bibinfo {author} {\bibfnamefont
  {T.}~\bibnamefont {Neupert}},\ }\href {\doibase 10.1038/s41567-018-0224-7}
  {\bibfield  {journal} {\bibinfo  {journal} {Nature Physics}\ }\textbf
  {\bibinfo {volume} {14}},\ \bibinfo {pages} {918} (\bibinfo {year}
  {2018})}\BibitemShut {NoStop}%
\bibitem [{\citenamefont {Hsu}\ \emph {et~al.}(2019)\citenamefont {Hsu},
  \citenamefont {Zhou}, \citenamefont {Chang}, \citenamefont {Ma},
  \citenamefont {Gedik}, \citenamefont {Bansil}, \citenamefont {Xu},
  \citenamefont {Lin},\ and\ \citenamefont {Fu}}]{bi_tci}%
  \BibitemOpen
  \bibfield  {author} {\bibinfo {author} {\bibfnamefont {C.-H.}\ \bibnamefont
  {Hsu}}, \bibinfo {author} {\bibfnamefont {X.}~\bibnamefont {Zhou}}, \bibinfo
  {author} {\bibfnamefont {T.-R.}\ \bibnamefont {Chang}}, \bibinfo {author}
  {\bibfnamefont {Q.}~\bibnamefont {Ma}}, \bibinfo {author} {\bibfnamefont
  {N.}~\bibnamefont {Gedik}}, \bibinfo {author} {\bibfnamefont
  {A.}~\bibnamefont {Bansil}}, \bibinfo {author} {\bibfnamefont {S.-Y.}\
  \bibnamefont {Xu}}, \bibinfo {author} {\bibfnamefont {H.}~\bibnamefont
  {Lin}}, \ and\ \bibinfo {author} {\bibfnamefont {L.}~\bibnamefont {Fu}},\
  }\href {\doibase 10.1073/pnas.1900527116} {\bibfield  {journal} {\bibinfo
  {journal} {Proceedings of the National Academy of Sciences}\ }\textbf
  {\bibinfo {volume} {116}},\ \bibinfo {pages} {13255} (\bibinfo {year}
  {2019})},\ \Eprint
  {http://arxiv.org/abs/https://www.pnas.org/content/116/27/13255.full.pdf}
  {https://www.pnas.org/content/116/27/13255.full.pdf} \BibitemShut {NoStop}%
\bibitem [{\citenamefont {Kruthoff}\ \emph {et~al.}(2019)\citenamefont
  {Kruthoff}, \citenamefont {de~Boer},\ and\ \citenamefont {van
  Wezel}}]{topology_trs}%
  \BibitemOpen
  \bibfield  {author} {\bibinfo {author} {\bibfnamefont {J.}~\bibnamefont
  {Kruthoff}}, \bibinfo {author} {\bibfnamefont {J.}~\bibnamefont {de~Boer}}, \
  and\ \bibinfo {author} {\bibfnamefont {J.}~\bibnamefont {van Wezel}},\ }\href
  {\doibase 10.1103/PhysRevB.100.075116} {\bibfield  {journal} {\bibinfo
  {journal} {Phys. Rev. B}\ }\textbf {\bibinfo {volume} {100}},\ \bibinfo
  {pages} {075116} (\bibinfo {year} {2019})}\BibitemShut {NoStop}%
\bibitem [{\citenamefont {Lau}\ \emph {et~al.}(2016)\citenamefont {Lau},
  \citenamefont {van~den Brink},\ and\ \citenamefont {Ortix}}]{1dtmi}%
  \BibitemOpen
  \bibfield  {author} {\bibinfo {author} {\bibfnamefont {A.}~\bibnamefont
  {Lau}}, \bibinfo {author} {\bibfnamefont {J.}~\bibnamefont {van~den Brink}},
  \ and\ \bibinfo {author} {\bibfnamefont {C.}~\bibnamefont {Ortix}},\ }\href
  {\doibase 10.1103/PhysRevB.94.165164} {\bibfield  {journal} {\bibinfo
  {journal} {Phys. Rev. B}\ }\textbf {\bibinfo {volume} {94}},\ \bibinfo
  {pages} {165164} (\bibinfo {year} {2016})}\BibitemShut {NoStop}%
\bibitem [{\citenamefont {Zhang}\ and\ \citenamefont
  {Murakami}(2021)}]{beyondtqc}%
  \BibitemOpen
  \bibfield  {author} {\bibinfo {author} {\bibfnamefont {T.}~\bibnamefont
  {Zhang}}\ and\ \bibinfo {author} {\bibfnamefont {S.}~\bibnamefont
  {Murakami}},\ }\href {\doibase 10.1088/1361-6463/ac13f4} {\bibfield
  {journal} {\bibinfo  {journal} {Journal of Physics D: Applied Physics}\
  }\textbf {\bibinfo {volume} {54}},\ \bibinfo {pages} {414002} (\bibinfo
  {year} {2021})}\BibitemShut {NoStop}%
\end{thebibliography}
%

\end{document}